\documentclass[useAMS,usenatbib]{mnras}
\usepackage{psfig,graphicx,amssymb,amsmath,rotating,nicefrac,color,natbib,tablefootnote}
\def\psq{$^{-2}$}

\def\lya{Ly$\alpha$}

\def\kps{\mathrm{km~s}^{-1}}

\def\arcsec{$^{\prime\prime}$}

\def\vdisp{\sqrt{\left<w_z^2\right>}}

\def\aj{AJ}%
\def\apj{ApJ}%
\def\apjl{ApJ}%
\def\apjs{ApJS}%
\def\ao{Appl.~Opt.}%
\def\aap{A\&A}%
\def\jcap{J. Cosmology Astropart. Phys.}%
\def\mnras{MNRAS}%
\def\nat{Nature}%

\defcitealias{2011MNRAS.414....2B}{Paper I}
\defcitealias{2011MNRAS.414...28C}{Paper II}
\defcitealias{2013MNRAS.430..425B}{Paper III}
\defcitealias{2014MNRAS.442.2094T}{Paper IV}

\def\ngal{1,665}
	
\def\tqso{29}

\def\nhrqso{16}

\title[VLT LBG Survey Redshift VI]{The VLT LBG Redshift Survey - VI. Mapping H{\sc i} in the proximity of $z\sim3$ LBGs with X-Shooter}
\author[R. M. Bielby et al.]{R. M. Bielby$^{1}$\thanks{E-mail:
richard.bielby@durham.ac.uk (RMB)} 
T. Shanks$^{1}$,
N. H. M. Crighton$^2$,
C. G. Bornancini$^{3,4}$,
\newauthor L. Infante$^{5}$, D. G. Lambas$^{3,4}$,
D. Minniti$^{6,7,8}$,
S. L. Morris$^1$,
P. Tummuangpak$^9$\\
$^1$ Durham University, South Road, Durham, DH1 3LE, United Kingdom\\
$^2$ Centre for Astrophysics and Supercomputing, Swinburne University of Technology, Hawthorn, Victoria 3122, Australia\\
$^3$ Instituto de Astronom\'{i}a Te\'{o}rica y Experimental (IATE, CONICET-UNC), Laprida 854, X5000BGR, C\'{o}rdoba, Argentina\\
$^4$ Observatorio Astron\'{o}mico (OAC), Universidad Nacional de C\'{o}rdoba, Laprida 854, X5000BGR, C\'{o}rdoba, Argentina\\
$^5$ Instituto de Astrof\'{i}sica, Facultad de F\'{i}sica, Centro de Astroingenneir\'{i}a, Pontificia Universidad Cat\'{o}lica de Chile\\ Av. Vicu\~{n}a Mackenna 4860, 782-0436 Macul, Santiago, Chile\\
$^6$ Departamento de Ciencias Fisicas, Universidad Andres Bello, Republica 220, Santiago, Chile\\
$^7$ Vatican Observatory, 00120 Vatican City State, Italy\\
$^8$ The Millennium Institute of Astrophysics (MAS), Santiago, Chile\\
$^9$ Department of Physics, Khon Kaen University, Khon Kaen, 40002, Thailand
}
\begin{document}

\date{}

\pagerange{\pageref{firstpage}--\pageref{lastpage}} \pubyear{2012}

\maketitle

\label{firstpage}

\begin{abstract}
We present an analysis of the spatial distribution of gas and galaxies using new X-Shooter observations of $z\sim3-4$ quasars. Adding the X-Shooter data to an existing dataset of high resolution quasar spectroscopy, we use a total sample of 29 quasars alongside $\sim1,700$ Lyman Break Galaxies (LBGs) in the redshift range $2\lesssim z\lesssim3.5$. Analysing the Ly$\alpha$ forest auto-correlation function using the full quasar sample, we find a clustering length of $s_0 = 0.081 \pm 0.006~h^{-1}$~Mpc. We then investigate the clustering and dynamics of Ly$\alpha$ forest absorbers around $z\sim3$ LBGs. From the redshift-space cross-correlation, we find a cross-clustering length of $s_0=0.27\pm0.14~h^{-1}$~Mpc, with power-law slope $\gamma=1.1\pm0.2$. We make a first analysis of the dependence of this clustering length on absorber strength based on cuts in the sightline transmitted flux, finding a clear preference for stronger absorption features to be more strongly clustered around the galaxy population than weaker absorption features. Further, we calculate the projected correlation function, finding a real-space clustering length of $r_0=0.24\pm0.04~h^{-1}$~Mpc (assuming a fixed slope $\gamma=1.1$). Taking this as the underlying real-space clustering, we fit the 2D cross-correlation function with a dynamical model incorporating the infall/redshift-space distortion parameter, $\beta_{\rm F}$,  and the peculiar velocity, $\vdisp$, finding $\beta_{\rm F}=1.02\pm0.22$ and $\vdisp=240\pm60$~km~s$^{-1}$. This result shows a significant detection of gas infall relative to the galaxy population, whilst the measured velocity dispersion is consistent with the velocity uncertainties on the galaxy redshifts. Finally, we evaluate the Cauchy-Schwarz inequality between the galaxy-galaxy, absorber-absorber, and galaxy-absorber correlation functions, finding a result significantly less than unity: $\xi_{\rm ag}^2/(\xi_{\rm gg}\xi_{\rm aa})=0.25\pm0.14$. This implies that galaxies and Ly$\alpha$ absorbers do not linearly trace the underlying dark matter distribution in the same way. 
\end{abstract}


\section{Introduction}

The relationship between gas and galaxies is a crucial component of galaxy formation models. Star-formation cannot be sustained without the supply of gas available to galaxies from their surroundings and understanding the flow of gas in and out of galaxies is imperative for a complete understanding of galaxy formation. Gas in the interstellar-medium (ISM) provides the reservoir from which the star-formation process is fuelled, but once this process is underway, winds from stars and supernovae begin to drive the interstellar gas outward \citep{Heckman90,1996ApJ...472..546L,wilman05}. The supernovae in particular may produce high-velocity winds which are able to act on large scales, driving gas and metals out of the galaxy and into the intergalactic medium, where $\approx80$\% of baryons reside at $z\sim3$ \citep{1997ApJ...479..523B}. Outflows are thought to heat and enrich gas within the galaxy halo, also referred to as the circum-galactic medium (CGM), reducing the amount of gas in the galaxy available for star-formation. A similar feedback process is expected to occur as a result of active galactic nuclei activity, driven by powerful jets. Simulations have indicated that these outflows may play a crucial role in galaxy evolution \citep[e.g.][]{springelhernquist03a,2005MNRAS.361..776S,2006MNRAS.370..645B,2006MNRAS.373.1265O,2007MNRAS.380..877S,2008MNRAS.387..577O,2010MNRAS.406.2325O,2010MNRAS.402.1536S,2010MNRAS.407.1581S,2011MNRAS.415...11D,2011MNRAS.417..950H,2012MNRAS.421.3522H,2014MNRAS.445..581H}.

Tracing the distribution and the dynamics of the gas presents a significant challenge to observational astronomy however. Although some progress has been made in recent years in attempting to trace the CGM gas via faint emission \citep[e.g.][]{2011ApJ...736..160S,2016A&A...587A..98W}, the prime method for probing both the CGM and inter-galactic medium (IGM) remains that of identifying absorption features in Quasi-Stellar Object (QSO) spectra. Given the ubiquity of hydrogen in the Universe, Lyman series Hydrogen absorption features offer insights into a range of environments, ranging across galaxy voids \citep[e.g.][]{2012MNRAS.425..245T}, filamentary structures \citep[e.g.][]{2016MNRAS.455.2662T}, and warm gas structures potentially tracing galaxy outflows or the intra-group medium \citep[e.g.][]{1994ApJ...427..696M,1999AJ....117..811H,2000ApJ...543L...9C,2016arXiv160105418P,2016arXiv160703386B}.

With the information from QSO sightline absorbers in hand, it is important to explore how these absorbers relate spatially to the galaxy population and large scale structure in general. Early statistical analyses of the distribution of absorbers with respect to the galaxy population showed tentative evidence for some clustering or `clumpiness' of absorbers around galaxies at $z\lesssim0.3$ \citep{1992ApJ...397...68B,1993ApJ...419..524M}. This early work was developed further, extending to larger samples and higher redshifts, by subsequent studies \citep[e.g.][]{1995ApJ...442..538L,1998ApJ...498...77C,2001ApJ...559..654C}. \citet{2006MNRAS.367.1261M} detected a significant correlation between H~{\sc i} absorbers and galaxies at separations of $\lesssim1.5$~Mpc, albeit weaker than the galaxy-galaxy auto-correlation. They found their results to be consistent with the absorbing gas and the galaxies coexisting in dark matter filaments and knots, consistent with predictions from models of galaxy formation.

At higher redshift, \citet{adelberger03} presented the first analysis of the cross-correlation between H{\sc i} in quasar sightlines and Lyman-break galaxies (LBGs) at $z\approx3$. Their results showed a clear increase in absorption in quasar sightlines within $\approx5~h^{-1}$~Mpc of the positions of LBGs in their survey. Further to this, at very small separations ($\lesssim500~h^{-1}$~kpc), they found a spike in the transmission profile suggestive of a lack of H{\sc i} gas in the immediate proximity of the $z\sim3$ galaxies. The authors suggested this may be the result of galaxy winds driving the gas away. However, the result at these scales was based on only 3 galaxies in their sample of $\sim800$ and it has proven difficult to recreate such a feature in simulations of star-forming galaxies. Following this work, the same group presented a similar analysis based on galaxies at $z\sim2$ \citep{adelberger05}. The same approach was taken using a larger sample of galaxies and no such spike in the absorption profile close to galaxies was observed, although some uncertainty remained given the differing redshift ranges of the two studies.

At $z\lesssim1$, further studies developed the low redshift absorber cross-correlation measurements \citep[e.g.][]{2006MNRAS.367.1251R,2007MNRAS.375..735W,2010MNRAS.402.2520S,2014MNRAS.437.2017T}. \citet{2006MNRAS.367.1251R} and \citet{2014MNRAS.437.2017T} are of particular interest in that they probe the 2D cross-correlation function, investigating evidence for dynamical effects on the absorber distribution around galaxies. Indeed two papers find somewhat conflicting results:\citet{2006MNRAS.367.1251R} showing prominent extensions in the correlation function along the line of sight (to $\approx400-600$~km~s$^{-1}$), claiming these to be the result of the intrinsic galaxy-absorber velocity dispersion; whilst \citet{2014MNRAS.437.2017T} place an upper limit of $\lesssim100$~km~s$^{-1}$ on the intrinsic velocity dispersion.

Following the same methods as \citet{adelberger03,adelberger05}, \citet[][hereafter Paper II]{2011MNRAS.414...28C} presented the first analysis of the H{\sc i}-galaxy cross-correlation using a wide area survey of LBGs at $z\sim3$ using the VIMOS instrument on the Very Large Telescope (VLT): namely the VLT LBG Redshift Survey (VLRS). This analysis was based on a sample of $\sim1,000$ $z\sim3$ galaxies surrounding 7 quasar sightlines at $z\sim3$. They found a deficit in Ly$\alpha$ transmission within $\sim5$~Mpc (comoving) in agreement with \citet{adelberger03} and \citet{adelberger05}, but no evidence for the upturn in average transmission seen by \citet{adelberger03}.

At a redshift range of $2<z<3$, \citet{2012ApJ...750...67R} and \citet{2012ApJ...751...94R} present analyses of the distribution of H{\sc i} gas around galaxies using $\sim800$ galaxies from the Keck Baryonic Structure Survey (KBSS). \citet{2012ApJ...750...67R} use Voigt profile fitted H{\sc i} absorption lines in 15 QSO spectra to analyse the gas distribution, finding evidence for infalling gas at large scales and peculiar velocities of $\approx260$~km~s$^{-1}$ at scales of $\lesssim400$ kpc. Similarly, \citet{2012ApJ...751...94R} use pixel-optical-depth (POD) analysis using the same data and reach equivalent conclusions. Although benefiting from high densities of galaxies, the KBSS fields are constrained to relatively small scales of $\approx6~h^{-1}$~Mpc (comoving) limiting the efficacy with which they may discern the Kaiser effect due to infalling material \citep{1987MNRAS.227....1K}.

Following this, \citet[][hereafter Paper IV]{2014MNRAS.442.2094T} presented the latest galaxy-H{\sc i} cross-correlation results of the VLRS survey, using an updated sample of 17 QSOs and $\approx2000$ $z\sim3$ LBGs. This built on the work of \citetalias{2011MNRAS.414...28C} again showing the flux decrement around LBGs and a lack of any transmission spike at small scales. \citetalias{2014MNRAS.442.2094T} presented a first full model-based analysis of redshift-space distortions (RSDs) of the gas around galaxies, finding a low velocity dispersion consistent with the redshift uncertainties on the galaxy positions and a tentative measurement of gas infall at large scales. Their observational results were presented alongside the analysis of a simulated volume from the Galaxies-Intergalactic Medium Interaction Calculation (GIMIC) hydrodynamical simulation which matched their observational data.

In this paper, we present an analysis of the relationship between gas and galaxies at redshifts of $z\approx3$ based on spectroscopic observations of Lyman-Break Galaxies (LBGs) and a combination of new moderate resolution quasar spectra from the VLT X-Shooter and high resolution quasar spectra. In section~2, we present the observations, detailing the data reduction for the X-Shooter data and giving an overview of the QSO and galaxy data used. Section~3 presents our analysis of the Ly$\alpha$ forest auto-correlation function incorporating the X-Shooter data, whilst in section~4 we present the galaxy-Ly$\alpha$ forest cross correlation analysis. We discuss our results in terms of the absorber dynamics and the relationship between absorbers and the underlying dark matter distribution in section~5. In section~6, we present our conclusions.

This is the sixth in a series of papers presenting the VLT LBG Redshift Survey (VLRS). \citet[][hereafter Paper I]{2011MNRAS.414....2B} presented the initial sample of $\sim1,000$ $z\approx3$ galaxies combined with an analysis of galaxy clustering. \citetalias{2011MNRAS.414...28C} analysed the gas-galaxy cross-correlation based on this first sample of $z\approx3$ galaxies. \citet[][hereafter Paper III]{2013MNRAS.430..425B} presented an updated sample totalling $\sim2,000$ LBGs and presented an analysis of galaxy clustering at $z\approx3$. \citetalias{2014MNRAS.442.2094T} built on the early VLRS results using GIMIC to analyse the results in the context of detailed simulations of galaxies and IGM. Finally, \citet[][Paper V]{2016MNRAS.456.4061B} presents a survey of Lyman-$\alpha$ emitters in a subset of the VLRS fields, although these data are not included in the analysis presented here (as we keep a focus on the relatively consistent selection of LBGs).

Throughout this paper, we use a cosmology given by $H_0=100~h~\kps$, $\Omega_m=0.3$, $\Omega_\Lambda=0.7$ and $\sigma_8=0.8$. Distances are quoted in comoving coordinates in units of $h^{-1}$~Mpc unless otherwise stated.

\section{Observations}

\subsection{Survey Overview}

The data used here covers 14 separate fields, each centred on a bright $z\gtrsim3$ QSO. Below we provide details of the QSO sightline and galaxy redshift data available in those fields.  

\subsubsection{High resolution QSO spectra}

\begin{figure}
\centering
\includegraphics[width=\columnwidth]{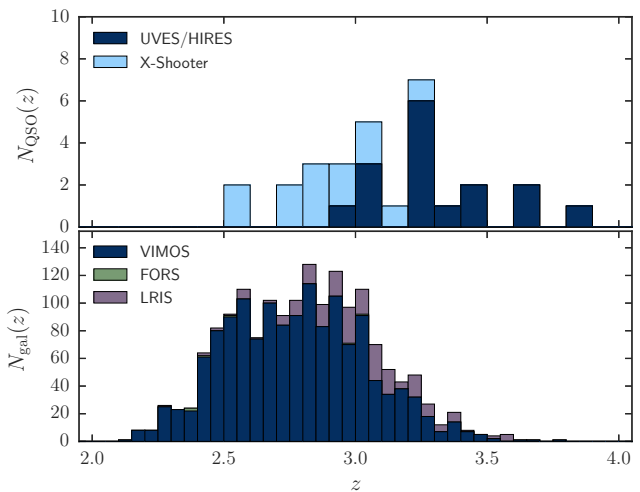}
\caption[]{Top panel: The stacked redshift distributions of the QSO sample observed with the high-resolution instruments HIRES and UVES (dark blue filled histogram) and the QSO sample observed with the moderate-resolution instrument X-Shooter (light blue filled histogram). Lower panel: The stacked redshift distribution of the \ngal\ spectroscopically observed galaxies included in this analysis (i.e. with redshifts lower than the redshift of any QSO in our sample of \tqso\ that lies within 30\arcsec), incorporating galaxies observed using VIMOS (dark blue histogram), FORS (green histogram) and LRIS (purple histogram).}
\label{fig:zdists}
\end{figure}

In the case of all but one of the fields, archive high resolution spectra are available from either VLT UVES or Keck HiRES for the central bright quasar (the exception being Q2359-0653). In addition, three further quasars ([WHO91] 0043-265, LBQS~0302-0019, and Q212904.90-160249.0) within the primary fields have archival high-resolution spectra available. In total, this provides \nhrqso\ QSOs with high resolution spectroscopy of the \lya\ forest. The redshift distribution of these is given by the dark blue histogram in Fig.~\ref{fig:zdists}.

We have performed the reduction of the available high-resolution Echelle spectra for these quasars and this is described in full in \citetalias{2011MNRAS.414...28C} and \citetalias{2014MNRAS.442.2094T}. An overview of these quasars is given in Table~\ref{tab:qso_info}.

For the purposes of our analysis in this paper, we wish to avoid the inclusion of Lyman Limit Systems (LLS) and Damped Ly$\alpha$ (DLA) systems (due to our pixel based analysis), as well as regions in the spectra that have not been observed. Taking individual features in the quasar spectra, the spectrum of Q0301-0035 is masked in the wavelength range $4460\mbox{\AA}<\lambda<4520\mbox{\AA}$, which contains a gap in the data. Also, the HIRES spectrum of J1201+0116 contains two gaps ($4420\mbox{\AA}<\lambda<4540\mbox{\AA}$ and $4790\mbox{\AA}<\lambda<4850\mbox{\AA}$), which are masked. The UVES spectrum of Q2348-011 contains a gap in the wavelength range $4510\mbox{\AA}<\lambda<4630\mbox{\AA}$, as well as DLAs at $\sim4160$\AA\ and $\sim4400$\AA. LLSs and DLAs are also present and masked in the spectra of WH091 0043-265 ($\lambda\sim4640$\AA), J0124+0044 ($\lambda\sim4950$\AA) and Q212904.90-160249.0 ($\lambda\sim4650$\AA).


\subsubsection{VLT X-Shooter spectra}

\citetalias{2011MNRAS.414...28C} presented a survey of $R<22$ quasars within the VLRS fields, with low resolution spectra of 295 quasars observed using AAOmega at the AAT. Here, we present VLT X-Shooter moderate resolution spectra of a selection of these, which overlap with the VLRS galaxy sample. In total we use 15 QSOs observed using the X-Shooter instrument as part of ESO programs 085.A-0327 and 087.A-0906 (2 of which also have high-resolution data available). The list of quasars used here is given in Table~\ref{tab:qso_info}, whilst the redshift distribution is shown by the pale blue component of the histogram in Fig.~\ref{fig:zdists}.

\begin{table*}
\centering
\caption{The full list of QSOs present in this study, observed with VLT UVES, VLT X-Shooter and Keck HIRES. In total we use 29 QSOs: 14 observed solely at high resolution, 13 observed solely at moderate resolution, and a further 2 observed at both high and moderate resolution. The resolution and S/N in the Ly$\alpha$ forest (Ly$\alpha$F) range are given for each QSO.
	}
\label{tab:qso_info}
\begin{tabular}{lllccccccc}
\hline
\hline
Name                     &  RA          & Dec         & $R$    & $z$  & Instrument    & Resolution & Ly$\alpha$F & $N_{\rm gal}$ & Galaxy \\
                         & \multicolumn{2}{c}{(J2000)}& (Vega) &      &               & (km~s$^{-1}$) & S/N    &  & data \\
\hline\hline
Q000033.06+070716.1      & 00:00:33.06 & $+$07:07:16.1 &  19.6  & 2.86  & X-Shooter   & 44         & 7.5 & 110   & V\\
Q000127.48+071911.8      & 00:01:27.48 & $+$07:19:11.8 &  20.7  & 2.87  & X-Shooter   & 44         & 3.4 & 110   & V  \\
Q000137.67+071412.2      & 00:01:37.67 & $+$07:14:12.2 &  20.8  & 2.99  & X-Shooter   & 44         & 5.6 & 109   & V  \\
Q2359+0653               & 00:01:40.60 & $+$07:09:54.0 &  18.4  & 3.23  & X-Shooter   & 44         & 28  & 97    & V  \\
Q000234.97+071349.3      & 00:02:34.97 & $+$07:13:49.3 &  20.6  & 2.60  & X-Shooter   & 44         & 5.0 & 50    & V  \\
\hline
LBQS 0041-2638           & 00:43:42.79 & $-$26:22:10.2 &  18.3  & 3.05  & X-Shooter   & 44         & 15.4& 165   & V\\
Q0042-2627               & 00:44:33.95 & $-$26:11:19.9 &  18.5  & 3.29  & HIRES       & 6.7        & 4.0 & 159   & V \\
$[$WHO91$]$ 0043-265     & 00:45:30.47 & $-$26:17:09.2 &  18.3  & 3.44  & HIRES       & 6.7        & 13.6& 139   & V \\
\hline
J0124+0044               & 01:24:03.78 & $+$00:44:32.7 &  17.9  & 3.81  & UVES        & 7.5        & 20.9& 72    & V\\
\hline
Q0201+1120               & 02:03:46.7  & $+$01:11:34.4 &  20.1  & 3.610 & HIRES       & 6.7        & 6.9 & 12    & L\\
\hline
Q0256-0000               & 02:59:05.6  & $+$01:00:11.2 &  18.2  & 3.364 & HIRES       & 6.7        & 28  & 37    & L\\
\hline
Q030241.61-002713.6      & 03:02:41.61 & $-$00:27:13.6 &  20.1  & 2.81  & X-Shooter   & 44         & 5.0 & 151   & V,L,F\\
SDSS J030335.42-002001.1 & 03:03:35.45 & $-$00:20:01.1 &  19.9  & 2.72  & X-Shooter   & 44         & 6.3 & 133   & V,L,F \\
LBQS 0301-0035           & 03:03:41.05 & $-$00:23:21.9 &  17.6  & 3.23  & HIRES       & 6.7        & 25  & 113   & V,L,F \\
LBQS 0302-0019           & 03:04:49.86 & $-$00:08:13.5 &  17.5  & 3.29  & HIRES       & 7.5        & 18  & 108   & V,L,F \\
\hline
QSO B0933+289            & 09:33:37.2  & +01:28:45.3   &  17.8  & 3.428 & HIRES       & 6.7        & 10  & 37    & L\\
\hline
HE0940-1050              & 09:42:53.50 & $-$11:04:25.9 &  16.6  & 3.05  & UVES        & 7.5        & 45  & 310   & V\\
\hline
J1201-0116               & 12:01:44.37 & $+$01:16:11.7 &  17.4  & 3.20  & HIRES       & 6.7        & 19  & 63    & V\\
\hline
Q1422+2309               & 14:24:38.1  & $+$01:22:56.0 &   .    & 3.620 & HIRES       & 6.7        & 59  & 69    & L\\
\hline
Q212904.90-160249.0      & 21:29:04.90 & $-$16:02:49.0 &  19.2  & 2.90  & UVES        & 6.7        & 6.7 & 89    & V\\
PKS2126-158              & 21:29:12.15 & $-$15:38:40.9 &  17.3  & 3.27  & UVES        & 7.5        & 72  & 126   & V \\
\hline
Q2231-0015               & 22:34:09.00 & $+$00:00:01.7 &  17.3  & 3.02  & UVES/X-Sh   & 7.5/44     & 40  & 80    & V,F\\
\hline
Q2233+136                & 22:36:27.2  & $+$01:13:57.1 &  18.7  & 3.210 & HIRES       & 6.7        & 8.0 & 29    & L\\
\hline
Q234919.94-010727.0      & 23:49:19.94 & $-$01:07:27.0 &  20.8  & 2.75  & X-Shooter   & 44         & 2.5 & 110   & V,F\\
SDSS J234921.56-005915.1 & 23:49:21.56 & $-$00:59:15.2 &  19.9  & 3.09  & X-Shooter   & 44         & 6.3 & 184   & V,F \\
Q234958.23-004426.4      & 23:49:58.23 & $-$00:44:26.4 &  21.0  & 2.58  & X-Shooter   & 44         & 2.9 & 78    & V,F \\
Q2348-011                & 23:50:57.88 & $-$00:52:09.9 &  18.7  & 3.01  & UVES/X-Sh   & 7.5/44     & 15  & 174   & V,F \\
Q235119.47-011229.2      & 23:51:19.47 & $-$01:12:29.2 &  20.1  & 2.94  & X-Shooter   & 44         & 4.6 & 157   & V,F \\
Q235201.36-011408.2      & 23:52:01.36 & $-$01:14:08.2 &  20.4  & 3.12  & X-Shooter   & 44         & 5.5 & 184   & V,F \\
\hline
\hline
\end{tabular}

\end{table*}

The observations were performed in NOD mode with individual exposure times of 694s, 695s and 246s with the UVB, VIS and NIR arms respectively. For quasars with magnitudes of $R\leq20$, 2 exposures were made in the UVB arm, 2 with the VIS arm and 6 with the NIR arm. Quasars fainter than $R=20$ were observed with double the number of exposures used for the brighter quasars. Slit widths of 1.0\arcsec, 1.2\arcsec and 1.2\arcsec were used for the UVB, VIS and NIR arms respectively, giving resolutions of $R=4,350$, $R=6,700$ and $R=3,890$ in each arm. Standard flux observations were made using the spectrophotometric standard stars GD71, LTT7987 and EG 131.

The X-Shooter spectra were reduced using the ESO X-Shooter pipeline package version number 1.4.6 and the {\sc esorex} command line reduction tool. We followed the standard reduction procedure as outlined in the X-Shooter Pipeline User Manual. All of the X-Shooter spectra were flux calibrated using the observed spectrophotometric stars.


\subsubsection{Quasar continuum fitting}
\label{sec:contfit}

In order to quantify the absorption along the line of sight to the observed quasars, we first need an estimate of the intrinsic quasar continuum and broad-line flux, $f_c(\lambda)$. This is performed using a suite of bespoke quasar absorption line python tools ({\sc qalpy}\footnote{Available online at https://github.com/nhmc/barak}) as developed and used by \citet{2010MNRAS.402.1273C,2010MNRAS.402.2520S} and \citetalias{2011MNRAS.414...28C} for the analysis of quasar spectra.

For each quasar, we use the script {\sc fitcontinuum} to estimate the intrinsic continuum, which follows the methods of \citet{1979ApJ...229..891Y} and \citet{1982MNRAS.198...91C}. Each QSO spectrum is first divided into wavelength intervals, in which the mean and standard deviation of the observed flux are calculated. A sigma-clipping process is then employed to iteratively reject the most aberrant pixels, until the remaining pixels show an approximately Gaussian distribution; the mean flux of these pixels is then taken to be the continuum level in that wavelength bin. An initial estimate of the continuum is then made by performing a cubic spline interpolation across the continuum levels estimated in each wavelength bin. Final adjustments are then made to the fit by hand where the continuum fit appears poor (i.e. around damped Ly$\alpha$ systems and emission lines). For further details of the script, we refer the reader to \citetalias{2011MNRAS.414...28C}.


\subsection{Galaxy data}

The galaxy data used in this analysis consists exclusively of galaxies selected using the Lyman Break method in optical filters ($U$, $B$ and $R$ or $u$, $g$ and $r$) and covering a redshift range of $2\lesssim z\lesssim3.5$. These cover a magnitude range of $23\lesssim R\lesssim25.5$. Whilst the full VLRS galaxy redshift data also incorporates Ly$\alpha$ emitters at $R\gtrsim25.5$ \citep{2016MNRAS.456.4061B}, we limit the sample in this paper to just the LBG selected population in order to use a consistently selected galaxy sample. In total, we use 1,665 galaxies lying at redshifts coeval with the Ly$\alpha$ forest of a background quasar. These galaxies are split between three observational sources: VLT VIMOS, VLT FORS and Keck LRIS. Each of these samples is described in further detail below. 

\subsubsection{VLT VIMOS data}

The current catalogue of LBGs from the VLRS has been presented in \citetalias{2013MNRAS.430..425B}. It consists of a sample of 2,147 spectroscopically confirmed galaxies, which were initially selected using the Lyman-break technique across 9 of the QSO fields discussed above. The galaxy sample was observed with low-resolution VIMOS spectroscopy using the LR\_Blue prism ($R\sim180$) under ESO programs 075.A-0683 (PI: D.~Minniti), 077.A-0612 (PI: T.~Shanks), 079.A-0442, 081.A-0418 and 082.A-0494 (PI: L.~Infante). Redshifts were measured based on fits to the Ly$\alpha$ and interstellar medium (ISM) spectral features, followed by corrections for the velocity offsets of these lines, giving velocity accuracies of $\sim360$~km~s$^{-1}$ \citepalias{2013MNRAS.430..425B}. Each galaxy identification was given a confidence flag from $0.5-1.0$, where \citetalias{2013MNRAS.430..425B} shows that $\sim60\%$ of $Q=0.5$ objects are correct based on multiple observations of single objects.

\begin{figure}
\centering
\includegraphics[width=80.mm]{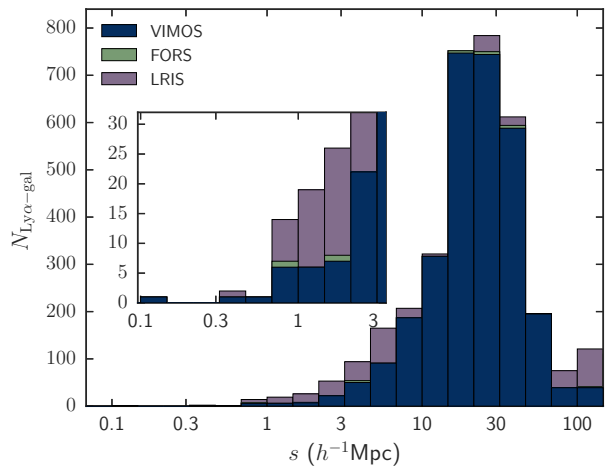}
\caption{Number of LBG-sightline pairs ($N_{\rm Ly\alpha-LBG}$) as a function of pair separation, $s$. The filled blue histogram shows the VIMOS detected galaxies, the green histogram the additional FORS detected galaxies, and purple the added LRIS galaxies (with the three histograms stacked to form the total). The inset shows a zoom in of the pair count histograms at small separation.}
\label{fig:paircounts_full}
\end{figure}

The numbers of VLRS $z\sim3$ galaxies in each field are given in Table~\ref{tab:qso_info}, whilst the redshift distribution is shown in Fig.~\ref{fig:zdists}. Taking the full sample of QSOs used in this study, we present the distribution of VLRS LBG - QSO sightline pairs as a function of separation, $s$ (in units of $h^{-1}$~Mpc), in Fig.~\ref{fig:paircounts_full} (blue histogram). The total number of available galaxies within the requisite redshift range (i.e. between the Ly$\alpha$ and Ly$\beta$ emission redshifts of background QSOs) to be included in our analysis is 1,437.

\subsubsection{VLT FORS data}

In addition to the VIMOS galaxy data, we also incorporate follow-up spectroscopy of galaxies close to quasar sight-lines using the VLT FORS instrument. These consist of re-observations of some galaxies to get more accurate redshifts plus observations of previously unobserved LBG candidates. The FORS observations were made close to three of the quasars used in the current study: Q0301-0035, Q2231-0015, and Q2348-011; and were observed as part of ESO program 087.A-0627.

The FORS data in these fields comprises reliable redshifts for 11 $z\sim3$ galaxies, 3 of which were originally identified using the VLRS VIMOS data and 8 of which lie within the requisite redshift range. These additional galaxies primarily lie $0.6<s<3h^{-1}$~Mpc from QSO sightlines in their respective fields, as shown in Fig.~\ref{fig:paircounts_full} (green histogram).

\subsubsection{Keck LRIS data}

Six of the fields are sampled by the Keck LRIS data of \citet{steidel03} and we incorporate these data into our sample. The data consists of 328 $z\sim3$ galaxy redshifts around the six available background QSOs, adding 220 additional galaxies within the Ly$\alpha$-Ly$\beta$ wavelength range of background QSOs to our total sample. The LRIS galaxy redshifts have been corrected for systematic shifts between features and have a velocity uncertainty of $\approx250$~km~s$^{-1}$.

We note that one of the \citet{steidel03} fields overlaps with one of the VLRS fields around the background QSO, LBQS~0302-0019. We find 3 galaxies observed by both surveys and find consistent redshifts in each case. For the purposes of this work, we use the Keck LRIS redshift in each of these three cases.

The Keck LRIS data are shown in the sightline-LBG pair counts in Fig.~\ref{fig:paircounts_full} (purple histogram) and predominantly add to the counts in the range $0.4<s<10~h^{-1}$~Mpc.

\section{Auto-correlation of the Ly$\alpha$-forest}

Using our \tqso\ moderate and high-resolution QSO spectra, we first measure the Ly$\alpha$-forest auto-correlation function. This constrains the mass distribution of the forest in preparation for then analysing the cross-correlation with the galaxy population. 

We perform the auto-correlation following the pixel to pixel method used in \citetalias{2011MNRAS.414...28C}. Using the normalised (transmitted) flux, $T=f/f_c$, we evaluate:

\begin{equation}
\delta_{\rm F}=\frac{T}{\overline{T}(z)}-1
\end{equation}

\noindent where $\overline{T}(z)$ is the mean normalised flux. If we discount high column density systems, the quantity $\delta_{\rm F}$ traces the underlying mass density fluctuations, $\delta_{\rm m}$, in a relatively linear manner \citep[e.g.][]{2002ApJ...580...42M}. The measured analytical form of $\overline{T}(z)$ at $z\approx3$ is given by the following \citep{2000ApJ...543....1M}:

\begin{equation}
\overline{T}(z) = 0.676 - 0.220(z-3)
\label{eq:tz}
\end{equation}

The auto-correlation function, $\xi_{\rm F}(r)$, is then given by:

\begin{equation}
\xi_{\rm F}(\Delta r) = \left<\delta_{\rm F}(r)\delta_{\rm F}(r+\Delta r)\right> 
\end{equation}

In this way, we calculate $\xi_{\rm F}$ using the individual normalised flux pixel values in each quasar sightline. To prevent the high-resolution data dominating our results in terms of sheer number of pixels, we resample the data (using a mean-binning) to match the X-Shooter pixel scale of 0.15~${\rm \AA}$/pixel.

\begin{figure}
\centering
\includegraphics[width=80.mm]{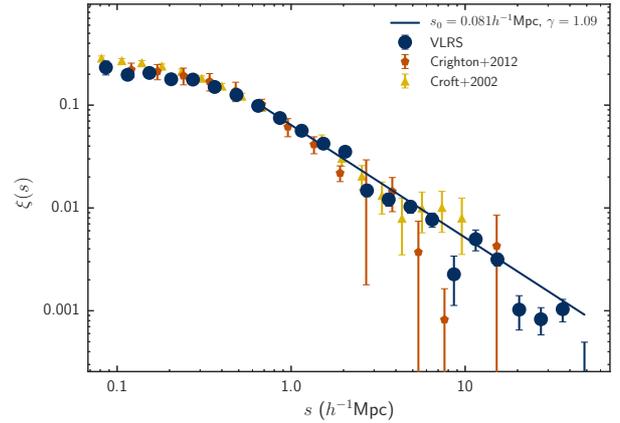}
\caption{The Ly$\alpha$ forest auto-correlation result for the 24 quasar sightlines studied here (dark filled circles). Pale filled squares show the result when limited to only the $\pi$-direction. The solid line shows the best fit to our full result at separations of $s>1.0~h^{-1}$~Mpc. Our previous results \citepalias{2011MNRAS.414...28C} are given by the pentagon points, whilst the literature results of \citet{2002ApJ...581...20C} are shown by filled triangles.}
\label{fig:xiforest}
\end{figure}

The result of the auto-correlation calculation using the \tqso\ quasar sightlines is shown in Fig.~\ref{fig:xiforest} (dark filled circles). This updated result is consistent with the previous VLRS results presented in \citetalias{2011MNRAS.414...28C} (orange pentagons) and \citetalias{2014MNRAS.442.2094T}, although we show the result to larger scales than in these previous papers. The use of moderate resolution data (and the necessary resampling of the high-resolution data to match) does not show any evidence of significantly affecting the results when compared to \citetalias{2011MNRAS.414...28C} (which is based on a subsample of our own data). The yellow triangles show the results of \citet{2002ApJ...581...20C}, where we take the average of their Subsamples D and E (see their Table~A6) in order to best match the mean QSO redshift of our sample. Again, the results are consistent where the scales probed overlap (and $s\gtrsim0.4~h^{-1}$Mpc) suggesting the measurements are robust.


\begin{figure}
	\centering
	\includegraphics[width=\columnwidth]{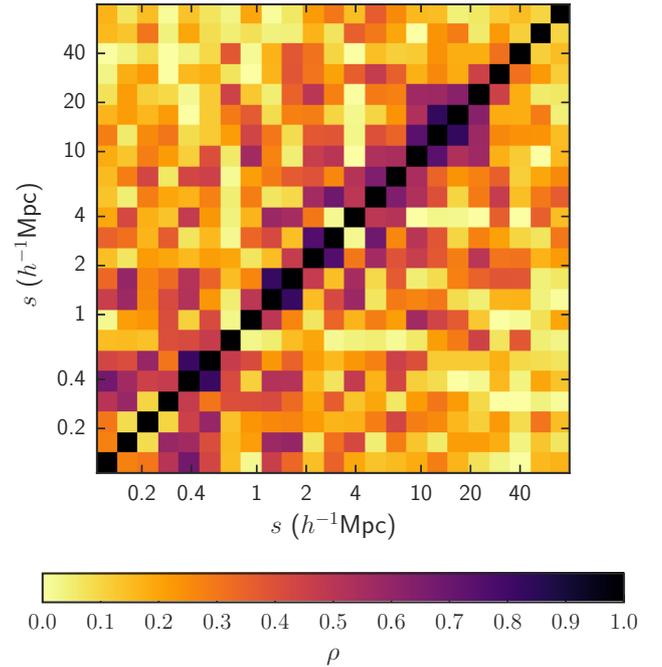}
	\caption{The covariance coefficient measured for the Ly$\alpha$ forest 1D auto-correlation shown in Fig.~\ref{fig:xiforest}}
	\label{fig:acorr_covariance}
\end{figure}

We evaluate the covariance of the auto-correlation function using the variance of the 29 individual QSO sightlines to evaluate the covariance coefficient, $\rho$ \citep[e.g.][]{wall2003practical}. The result is shown in Fig.~\ref{fig:acorr_covariance}. Strong covariance between adjacent bins is clearly seen with values of $\rho\gtrsim0.7$. The covariance coefficient generally falls to values of $\rho\lesssim0.5$ for bins separated by $\sim2$ or more bins. 

Fitting a simple power-law function of the form $\xi_{\rm F}(s)=(s/s_0)^{-\gamma}$ to our result at scales of $s>0.9$~Mpc, we find a best fit correlation length of $s_0=0.081\pm0.006~h^{-1}$Mpc and a slope of $\gamma=1.09\pm0.04$. The amplitude of the Ly$\alpha$ forest clustering is significantly less than the clustering of the galaxy population at $z\sim3$ (the corresponding galaxy correlation length is $s_0\sim3-5~h^{-1}$~Mpc, e.g. \citetalias{2013MNRAS.430..425B}).

\section{Distribution of Hi around LBGs}

\subsection{\lya--LBG cross-correlation}

We calculate the LBG-Ly$\alpha$ cross-correlation using the following commonly used form \citep[see for example][]{1973ApJ...185..413P,1979A&A....74..308S}:

\begin{equation}
\xi(s) = \frac{\left<D_{\rm g}D_{\rm T}(s)\right>}{\left<D_{\rm g}R_{\rm T}(s)\right>}-1
\end{equation}

\noindent where $\left<D_{\rm g}D_{\rm T}(s)\right>$ is the weighted galaxy-pixel pair-count at a given separation, $s$, and $\left<D_{\rm g}R_{\rm T}(s)\right>$ is the weighted galaxy-random-pixel pair-count as a function of separation. The weighting of a given pair is given by the flux transmission, adjusted for redshift, such that the weighted pair-count is given by $\left<D_{\rm g}D_{\rm T}(s)\right>=\Sigma\left(T(s)/\overline{T}(z)\right)$.

Damped \lya\ absorbers (DLAs), systems with hydrogen column densities $>10^{20}$ cm\psq, produce \lya\ absorption lines with highly broadened wings. It is therefore important to mask DLA absorption features out of the spectra, since the broad wings would introduce a significant bias when inferring the neutral hydrogen distribution along the sightline from the measured \lya\ transmitted flux. We note also that we only use the quasar spectral range between Ly$\beta$ and \lya, since shortward of Ly$\beta$ we may not effectively isolate the \lya\ forest from the Ly$\beta$ forest.

Using the method outlined here, we use our sample of QSO spectra to infer the gas density of the intergalactic medium along the quasar sightlines. By cross-correlating this information with the positions of our LBG sample, we can attempt to measure the interactions between galaxies and the IGM at high redshift.

\begin{figure}
\centering
\includegraphics[width=80.mm]{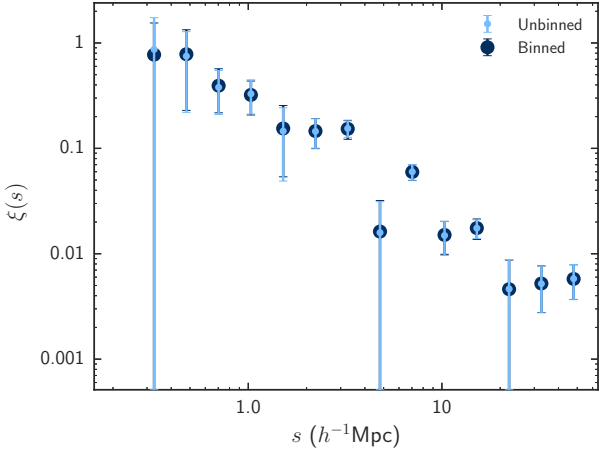}
\caption{The result of the cross-correlation analysis using only the 11 background quasars from VLRS with high-resolution spectra. The large circular points show the result based on the native pixel scale of the high resolution data (0.03~\AA/pixel), whilst the small circles show the data re-sampled to the X-Shooter pixel scale (0.15~\AA/pixel).}
\label{fig:hiresxcorr}
\end{figure}

Given the resampling of the high-resolution data, we verify that this has minimal effect on our results by performing the cross-correlation with only the high-resolution QSO spectra, using the non-resampled and the resampled data-sets. The result is shown in Fig.~\ref{fig:hiresxcorr}. Small deviations are observed in the binned sample at separations of $\lesssim5~h^{-1}$~Mpc, however these changes from the unbinned sample are much smaller than the estimated statistical uncertainties on the points themselves.


\begin{figure}
\centering
\includegraphics[width=80.mm]{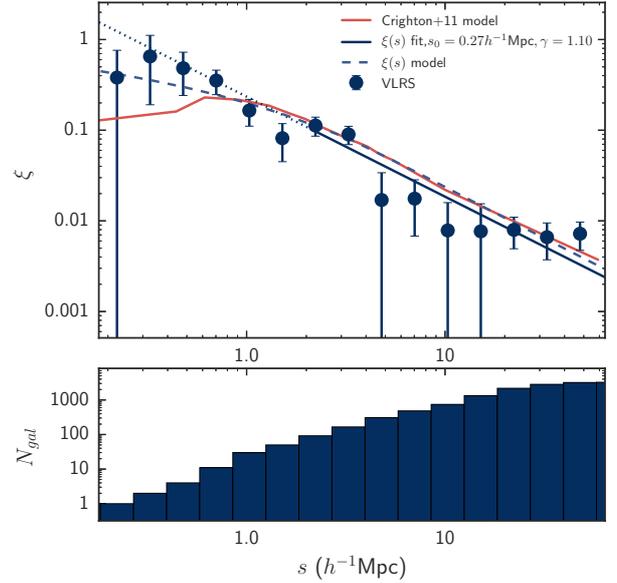}
\caption{The result of the cross-correlation analysis using the \tqso\ background quasars (filled circles). The solid blue curve shows the best fitting power-law to the data, whilst the dashed curve shows the $\xi(s)$ model derived from fitting the 2D cross-correlation in Section~\ref{sec:2dccf}. The red curve shows a model taken from \citetalias{2011MNRAS.414...28C} with a $0.5~h^{-1}$~Mpc `transmission spike' in the Ly$\alpha$ transmission profile around star-forming galaxies (and smoothed by a velocity dispersion of 150~km~s$^{-1}$). In the lower panel, we show the numbers of galaxy-absorber pairs as a function of separation, $s$.}
\label{fig:fullxcorr}
\end{figure}

\begin{figure}
	\centering
	\includegraphics[width=80.mm]{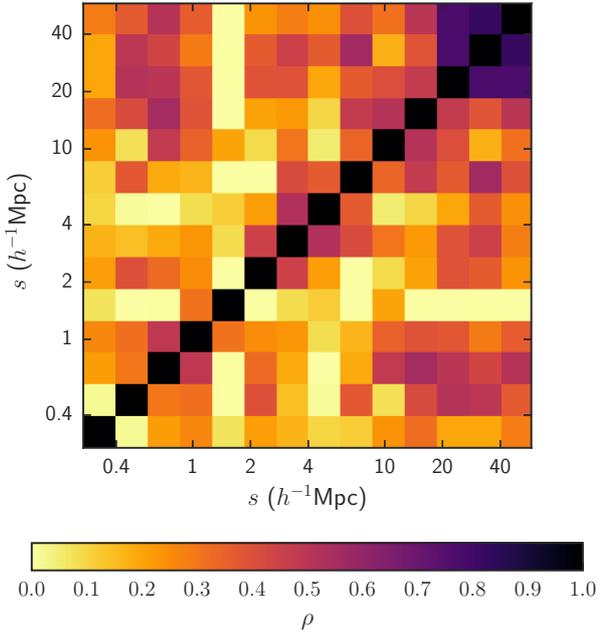}
	\caption{The covariance coefficient measured for the 1D cross-correlation shown in Fig.~\ref{fig:fullxcorr}}
	\label{fig:xis_covariance}
\end{figure}

The result of the cross-correlation with the \tqso\ QSO spectra is shown in Fig.~\ref{fig:fullxcorr} (filled circles) and follows a power-law form. This measurement of the cross-correlation will inevitably contain some covariance between individual bins, as with all such correlation functions. As with the Ly$\alpha$ forest auto-correlation, we again evaluate the covariance coefficient, $\rho$, using the individual $\xi(s)$ measurement for each of the 29 background quasars. The result is shown in Fig.~\ref{fig:xis_covariance}. We find a covariance coefficient of $\rho\lesssim0.5$, except for at the high separation bins (i.e. $s\gtrsim16$), which are strongly correlated with each other ($\rho\approx0.8$).

Following the same method as we applied for the auto-correlation, we fit a power-law to the cross-correlation of the form $\xi(s)=\left(s/s_0\right)^{-\gamma}$, finding a correlation length of $s_0=0.27\pm0.14~h^{-1}$~Mpc and a slope of $\gamma=1.1\pm0.2$ (blue curve in Fig.~\ref{fig:fullxcorr}). For reference, we also plot in Fig.~\ref{fig:fullxcorr} the transmission spike model of \citetalias{2011MNRAS.414...28C} (red curve) incorporating a power law function ($s_0=0.3~h^{-1}$~Mpc and $\gamma=1.0$) with a transmission spike of width $0.5~h^{-1}$~Mpc, convolved with a velocity dispersion of 150~km~s$^{-1}$. As with \citet{adelberger05}, \citetalias{2011MNRAS.414...28C, 2014MNRAS.442.2094T} and \citet{2012ApJ...751...94R}, the data shows no evidence for a transmission  spike \citep{adelberger03}.

\begin{figure}
\centering
\includegraphics[width=80.mm]{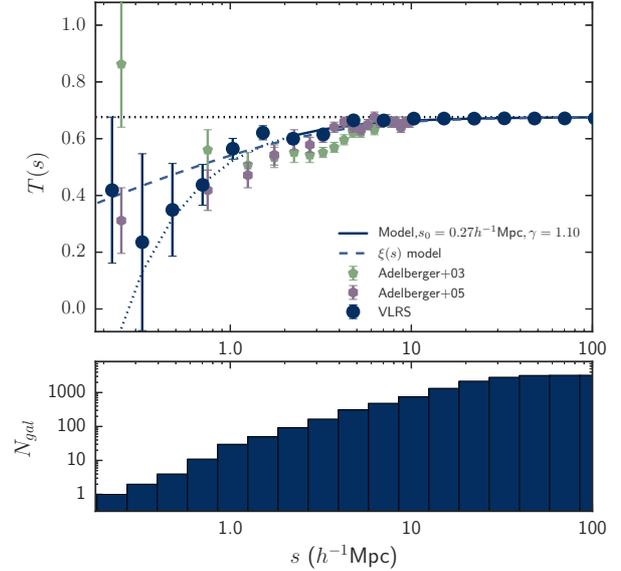}
\caption{The result of the cross-correlation analysis weighted by transmitted flux based on the \tqso\ background quasars from VLRS with moderate and high-resolution spectra (filled circles). The power law fit to the cross-correlation (also shown in Fig.~\ref{fig:fullxcorr}) is again shown by the dashed/solid line, whilst the dashed curve again shows the result from the 2D cross-correlation fitting presented in Section~\ref{sec:2dccf}. Also shown are the results of \citet[][filled pentagons]{adelberger03} and \citet[][filled hexagons]{adelberger05}. The lower panel shows the numbers of galaxy-QSO pairs used in each bin for our analysis.}
\label{fig:fulltrans}
\end{figure}

For the purposes of comparison with previous works, we present the cross-correlation in terms of the flux transmission, $\left<T(s)\right>=(1-\xi(s))\overline{T}(z=3)$, in Fig.~\ref{fig:fulltrans}. Again our results are shown by the large blue filled circles. For comparison, the results of \citet[][filled pentagons]{adelberger03} and \citet[][filled hexagons]{adelberger05} are also shown. Consistent with the \citet{adelberger05} result at $z\approx2.4$, our data shows strong absorption within $\approx5~h^{-1}$~Mpc of the galaxy population. Although reduced absorption is seen in the closest spatial bin, the data point is consistent with monotonically increasing absorption.

There are few previous measurements of the cross-correlation clustering length of galaxies and the Ly$\alpha$ forest, and at redshift $z\approx3$ we are limited to comparing with our own work in this respect. \citetalias{2011MNRAS.414...28C} showed their data to be consistent with a power law given by $s_0=0.3$ and $\gamma=1.0$, which is consistent to within $\approx1\sigma$ with this latest result. Motivated in part by simulation results, \citetalias{2014MNRAS.442.2094T} attempted to fit a double power-law to the cross-correlation, finding $s_0=0.49\pm0.32$ and $\gamma=1.47\pm0.91$ at scales of $s\gtrsim1~h^{-1}$~Mpc, and $s_0=0.08\pm0.04$ and $\gamma=0.49\pm0.32$ at smaller scales. We find instead that our results are well fit by a single power-law, leading to a lower overall clustering length and shallower slope than found using a double power law as in \citetalias{2014MNRAS.442.2094T}. We note however, that given the large quoted errors on the \citetalias{2014MNRAS.442.2094T} result, our single power-law result is in fact consistent with their $s\gtrsim1~h^{-1}$~Mpc result. The present more accurate value should be preferred.

Looking to lower redshift, comparisons are made more complex in that measurements are more commonly made using Voigt profile fitting individual lines as opposed to the pixel-based method used here. How these two differently constructed measurements relate to each other has not been fully investigated, however for completeness we note here the results of \citet{2014MNRAS.437.2017T} who performed an analysis of the cross-correlation of galaxies and H~{\sc i} absorption features in QSO sightlines in the redshift range $0<z<1$. For their full sample of identified absorbers (with a column density range of $10^{13}<N_{\rm HI}<10^{17}$~cm$^{-2}$) they find a correlation length of $r_0=1.12\pm0.14~h^{-1}$~Mpc (with a slope of $\gamma=1.4\pm0.1$). On the other hand for a low-column density subset ($10^{13}<N_{\rm HI}<10^{14}$~cm$^{-2}$), they determine a correlation length of $r_0=0.14\pm0.28~h^{-1}$~Mpc. Our own measurement (which is based on masking large absorbers from the sightlines) is consistent with the lower column density range of \citet{2014MNRAS.437.2017T}.

\subsection{The effect of absorber strength on the cross-correlation}

Prompted by \citet{2014MNRAS.437.2017T}, we now investigate the presence of any dependence of our results on absorber strength. Whilst the general Ly$\alpha$ forest traces the IGM, as we probe stronger absorbers, current models suggest that we are more likely to be tracing gas structures associated with galactic halos and filaments. It is interesting therefore to analyse the cross-correlation function as a function of limiting transmitted flux. Indeed, we can relate the measured pixel transmitted flux to the column density of absorbers as discussed by \citet{2014MNRAS.441.1718P}. Taking the values directly from Table~1 in \citet{2014MNRAS.441.1718P}, $T_{\rm lim}<0.25$ equates approximately to a column density of $N_{\rm HI}=16.5$~cm$^{-2}$, whilst a cut of $T_{\rm lim}<0.45$ corresponds to $N_{\rm HI}=14.5$~cm$^{-2}$.

\begin{figure}
\centering
\includegraphics[width=80.mm]{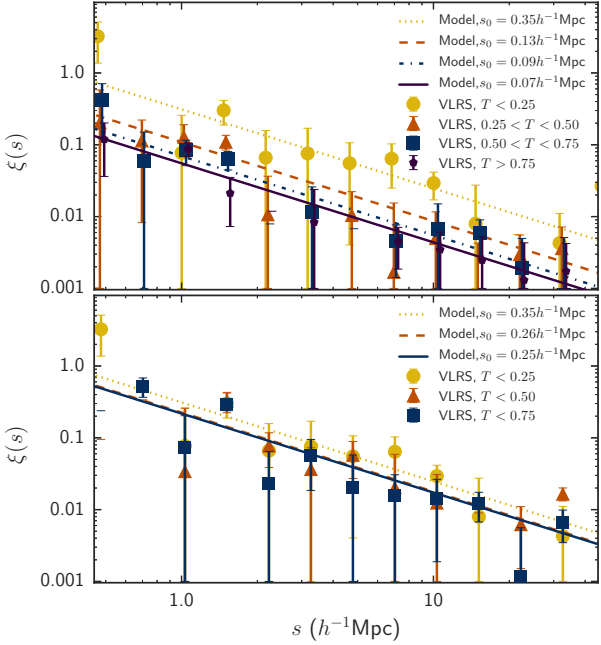}
\caption{\emph{Top-panel}: Cross-correlation results as a function of transmitted flux bins, with results given for $T>0.75$ (pentagons), $0.50<T<0.75$ (dark blue squares), $0.25<T<0.50$ (orange triangles) and $T<0.25$ (gold circles). Power law fits are presented for each with the slope fixed to $\gamma=1.2$. \emph{Lower-panel}: The quivalent results as a function now as a function of a cut in transmitted flux, with results given for $T<0.75$ (dark blue squares), $T<0.50$ (orange triangles) and $T<0.25$ (gold circles). Power law fits are again presented for each with the slope fixed to $\gamma=1.2$ in all three cases. The solid, dashed and dotted lines show the fits to the $T<0.75$, $T<0.50$ and $T<0.25$ results respectively.}
\label{fig:xcorr_flim}
\end{figure}

We have performed our cross-correlation analysis with the sightline normalised flux pixels grouped by both limiting normalised flux and discrete bins in normalised flux. The results using the sightline pixels in flux bins of $T_{\rm lim}\leq0.25$ ($N_{\rm HI}\gtrsim16.5$), $0.25\leq T_{\rm lim}\leq0.50$ ($N_{\rm HI}\approx14-16.5$), $0.50\leq T_{\rm lim}\leq0.75$ and $T_{\rm lim}\geq0.75$ are shown in the top panel of Fig.~\ref{fig:xcorr_flim}, whilst the results using upper limits of $T_{\rm lim}\leq0.25$ ($N_{\rm HI}\gtrsim16.5$), $T_{\rm lim}\leq0.50$ ($N_{\rm HI}\gtrsim14$), and $T_{\rm lim}\leq0.75$ are shown in the lower panel. Some dependence of the clustering amplitude is evident in both sets of results, with lower normalised flux (high column density) corresponding to higher amplitude.

\begin{figure}
\centering
\includegraphics[width=80.mm]{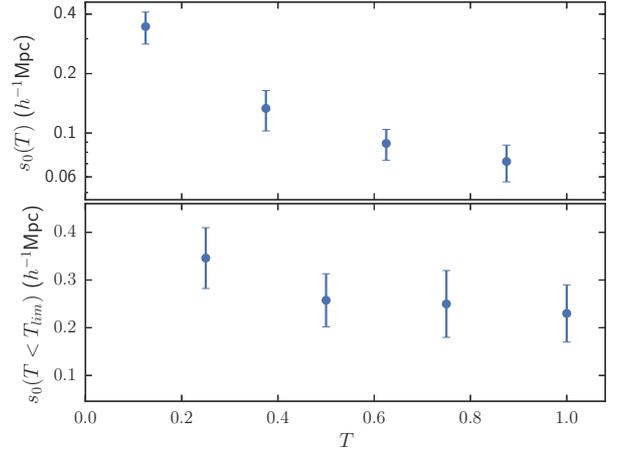}
\caption{Measured values of the clustering length, $s_0$, as a function of flux binning (top panel) and a cut in the Ly$\alpha$ transmission flux, $T_{\rm lim}$ (lower pabel). The values correspond to the power-law fits shown in Fig.~\ref{fig:fullxcorr} and Fig.~\ref{fig:xcorr_flim}. The data show evidence for a dependence of the clustering length on the cut in Ly$\alpha$ transmission flux, such that stronger absorption systems are more highly correlated with galaxies at $z\sim3$.}
\label{fig:xcorr_s0vTlim}
\end{figure}

We fit the samples with the power-law form already used, taking a fixed power-law slope of $\gamma=1.1$. The resulting power-laws are shown in Fig.~\ref{fig:xcorr_flim} as described in the figure legends. We show the resulting clustering lengths, $s_0$, for both the discrete bins (top-panel) and flux limited bins (bottom panel) in Fig.~\ref{fig:xcorr_s0vTlim}. A trend is observed with the clustering length decreasing with increasing normalised flux in both panels, with the discrete binned sample showing the clearest trend (at the $\approx4\sigma$ level between the minimum and maximum values).

Significant progress has been made with low-redshift studies of the relationship between galaxies and the Ly$\alpha$ forest in recent years with the advent of the Cosmic Origins Spectrograph (COS). Of direct relevance to our high-redshift studies are the results of \citet{2014MNRAS.437.2017T}, who calculated the projected and two-dimensional cross-correlation functions between Ly$\alpha$ absorbers and galaxies. Given the lower rate of Ly$\alpha$ absorber blending in the forest at lower redshift \citet{2014MNRAS.437.2017T} were able to conduct a complete Voigt profile analysis and perform the cross-correlation using individual absorbers as opposed to using the binned normalised fluxes as we do here. Given their Voigt profile analysis, they are able to cover a much wider range in column density than presented here with our pixel based method, finding $r_0=0.2\pm0.4~h^{-1}$~Mpc for $N<10^{14}$~cm$^{-2}$ and $r_0=3.8\pm0.2~h^{-1}$~Mpc for $N\geq10^{14}$~cm$^{-2}$ (using their star-forming galaxy sample).

\subsection{The projected and 2D cross-correlation function}
\label{sec:2dccf}
In order to isolate and analyse the effects of gas and galaxy dynamics on the cross-correlation, we now turn to the projected and 2D LBG-Ly$\alpha$ cross-correlation functions.

Following \citetalias{2014MNRAS.442.2094T}, we calculate the 2D cross-correlation function, $\xi(\sigma,\pi)$, and project this along the $\pi$ direction to derive the projected correlation function, $w_{\rm p}(\sigma)$. We calculate $\xi(\sigma,\pi)$ identically to $\xi(s)$ but now on a two-dimensional grid of $\sigma$ (on-sky separation) and $\pi$ (line-of-sight separation). The projection to $w_{\rm p}(\sigma)$ is performed by integrating $\xi(\sigma,\pi)$ to some limit in $\pi$:

\begin{equation}
w_{\rm p} = 2\int^{\pi_{\rm max}}_0 \xi(\sigma,\pi) {\rm d}\pi
\end{equation}

As in the $\xi(s)$ case, we calculate uncertainties on our $\xi(\sigma,\pi)$ and $w_{\rm p}(\sigma)$ measurements by performing the analysis on 50 galaxy catalogues each taking the same on-sky spatial distribution as the data, but with randomised redshifts using the measured galaxy redshift distribution as the probability density function.

\begin{figure}
\centering
\includegraphics[width=\columnwidth]{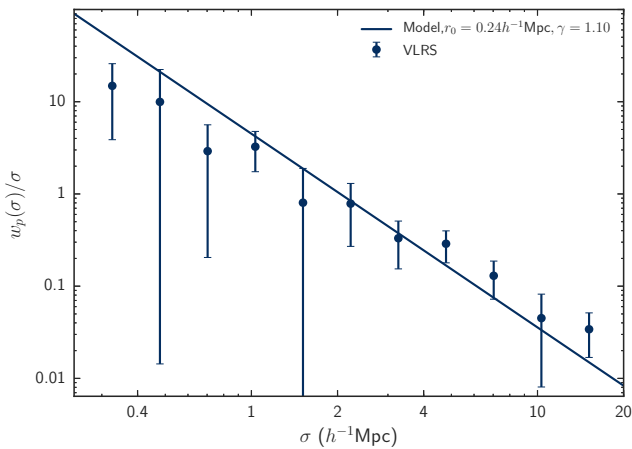}
\caption{The projected cross correlation function obtained from the 2D cross-correlation map (shown in Fig.~\ref{fig:full2dxcorr}) for the sample of 29 QSO sightlines (solid circle points). The solid line shows a power law fit to the data (at $1~h^{-1}~{\rm Mpc}<\sigma<20~h^{-1}~{\rm Mpc}$), given by $r_0=0.24\pm0.04~h^{-1}$~Mpc and a fixed slope of $\gamma=1.1$.}
\label{fig:fullwpsig}
\end{figure}

The resulting projected correlation function is shown in Fig.~\ref{fig:fullwpsig}. Again we fit a power law of the form $\xi(r)=(r/r_0)^{-\gamma}$ to the data, excluding separations of $\sigma<1~h^{-1}$~Mpc (to avoid small scale effects due to line saturation and line broadening), whilst fitting the data up to a scale of $\sigma<20~h^{-1}$~Mpc. This gives a best fit clustering length of $r_0=0.24\pm0.04~h^{-1}$~Mpc (based on a fixed slope of $\gamma=1.1$ for consistency with the $\xi(s)$ measurement).

\begin{figure*}
\centering
\includegraphics[width=\textwidth]{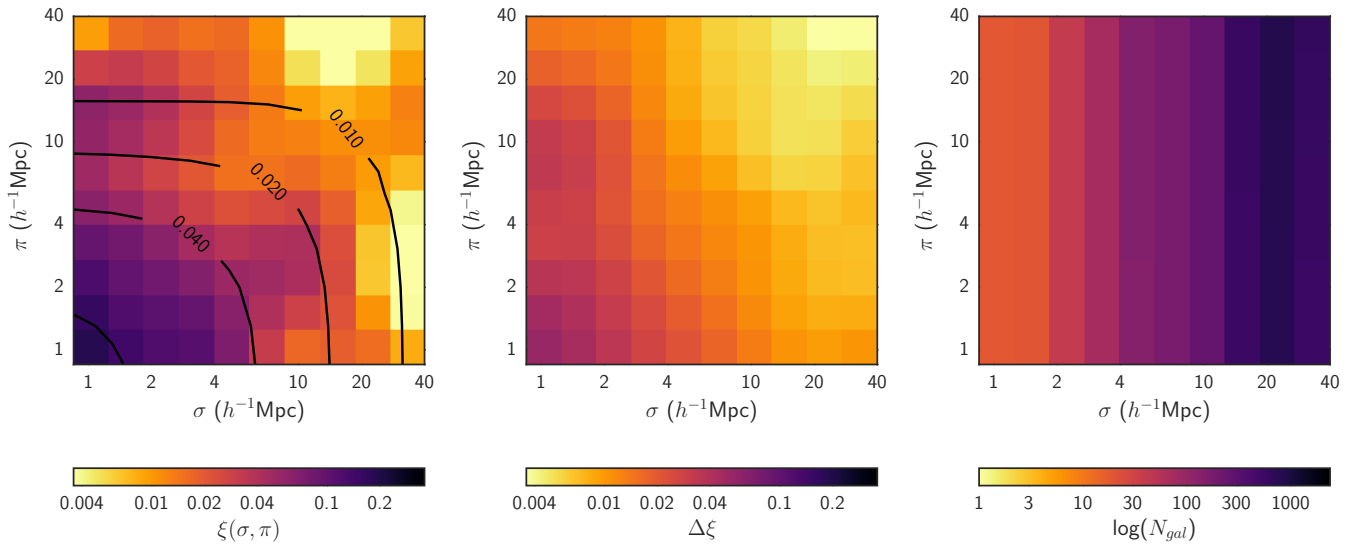}
\caption{\emph{Left}: The $z\approx3$ 2D galaxy galaxy-Ly$\alpha$ cross correlation function, $\xi(\sigma,\pi)$ for the full (VLT+Keck) sample (shaded contour map). The solid contours show the best fitting RSD model with $\beta_{\rm F}=1.02\pm0.22$ and $\vdisp=240\pm60$. \emph{Centre}: The estimated uncertainties on $\xi(\sigma,\pi)$, calculated based on 50 random realisations of the galaxy catalogue. \emph{Right}: The number of galaxies used in each bin.}
\label{fig:full2dxcorr}
\end{figure*}

The 2D cross-correlation result for our sample of galaxies and QSO sightlines is given in Fig.~\ref{fig:full2dxcorr}. The filled contour map in the left hand panel shows the 2D cross-correlation measurement, $\xi(\sigma,\pi)$, the central panel shows the calculated $1\sigma$ uncertainties on the measurement, and the right-hand panel shows the number of galaxy-sightline pairs in each bin.

We now proceed to extract the velocity field information from the $\xi(\sigma,\pi)$ measurement. For this we use the velocity field model discussed in \citetalias{2014MNRAS.442.2094T}, which incorporates the effects of small scale velocity dispersion, characterised by the parameter $\vdisp$, and large scale infall velocity fields, characterised by the infall parameter $\beta_{\rm F}$ (also known as the redshift-space distortion parameter). For a detailed explanation and discussion of the model and these parameters, we refer the reader to \citetalias{2014MNRAS.442.2094T} and \citet{hawkins03}.

For the purposes of the fitting of the velocity field parameters, we fix the underlying real-space correlation-function to that derived from $w_{\rm p}(\sigma)$, i.e. a power law with correlation length $r_0=0.24~h^{-1}$~Mpc and a slope of $\gamma=1.1$. Further, we fix the galaxy infall parameter, $\beta_{\rm gal}$, to the value derived for the LBG sample in \citetalias{2013MNRAS.430..425B}: $\beta_{\rm gal}=0.36$.

\begin{figure}
\centering
\includegraphics[width=\columnwidth]{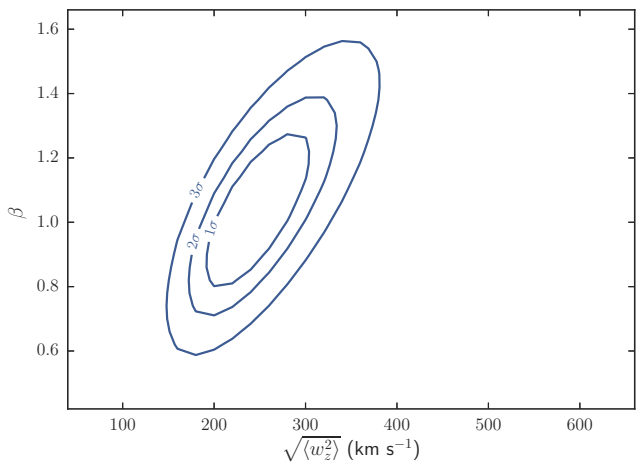}
\caption{Confidence contours on fitting the 2D cross-correlation function for the infall parameter, $\beta$ and the velocity dispersion, $\vdisp$, for the combined VLT$+$Keck data sample. The fit assumes an underlying power law form to the intrinsic cross-correlation function with correlation length $r_0=0.24~h^{-1}$~Mpc and a slope of $\gamma=1.1$.} 
\label{fig:full2dxcorrfit}
\end{figure}

The resulting measurements of $\vdisp$ and $\beta_{\rm F}$, derived from a $\chi^2$ analysis, are shown in Fig.~\ref{fig:full2dxcorrfit}. We show the confidence limits on the best fitting values for the infall parameter, $\beta_{\rm F}=1.02\pm0.22$, and the velocity dispersion, $\vdisp=240\pm60$~km~s$^{-1}$. This best fitting model is shown in the left hand panel of Fig.~\ref{fig:full2dxcorr} by the contour lines.

\section{Discussion}

\subsection{Absorber dynamics}

\subsubsection{Infall/RSD parameter}

Few statistical or quantitative measures of the Ly$\alpha$ forest absorber-galaxy dynamics have been made at any redshift, whilst model predictions are also relatively sparse. The primary route to constraining the dynamics of the Ly$\alpha$ forest has instead been via the Ly$\alpha$ forest auto-correlation function or power spectrum \citep[e.g.][]{2003ApJ...585...34M, 2011JCAP...09..001S,2012JCAP...03..004S}. \citet{2013MNRAS.433.3103R} showed evidence of large scale infall in their cross-correlation analysis, but were limited to scales of $\lesssim7~h^{-1}$~Mpc and did not present any constraints on the large scale dynamics themselves.

\citetalias{2014MNRAS.442.2094T} presented an analysis of these dynamics with both a subset of the data used here and a sub-volume snapshot of the GIMIC simulation. Using equivalent analyses to those presented here, the data gave constraints on the infall parameter of $\beta_{\rm F}=0.33^{+0.23}_{-0.33}$, whilst the analysis of the simulation predicted a somewhat higher value of $\beta_{\rm F}=0.51\pm0.12$ (although the authors note that this simulated measurement may be affected by the relatively small size of the simulated volume). We have now improved significantly on the previous result by adding the X-Shooter data to the analysis, in particular improving our constraints on the underlying input (real-space) model given by the $w_p(\sigma$) measurement. Given the improved constraints on $r_0$ from $w_p(\sigma)$ and the enhanced signal on the 2D cross-correlation signal, we find that our measurement of the infall parameter is higher than suggested in \citetalias{2014MNRAS.442.2094T}.

Measurement of the large scale dynamics of the Ly$\alpha$ forest from the auto-correlation function has largely been based on BOSS, with original results from BOSS giving constraints of $\beta=0.8\pm0.2$ \citep{2011JCAP...09..001S}. Using the Ly$\alpha$ forest measurements of BOSS Data Release 11 ($z=2.3$), \citet{2015A&A...574A..59D} and \citet{2015JCAP...11..034B} find best fitting measurements of $\beta_{\rm F}=1.50\pm0.47$ and $\beta_{\rm F}=1.39^{+0.11}_{-0.10}$ respectively. Closer in method to our own work, \citet{2013JCAP...05..018F} derive $\beta_{\rm F}$ from the cross-correlation between the Ly$\alpha$ forest and the BOSS QSO sample at $z\approx2.3$, finding $\beta_{\rm F}=1.10^{+0.17}_{-0.15}$. Most models predict some evolution in the infall parameter between the BOSS redshift and that of our analysis. Indeed, the simulation results of \citet{2015JCAP...12..017A} would suggest a factor of $\beta_{\rm F}(z=2.8)/\beta_{\rm F}(z=2.3)\approx0.92$, translating the most recent BOSS results to $\beta_{\rm F}\approx1.28-1.38$ at $z=2.8$. Our own result lies somewhere between the early and more recent BOSS results and, at $\beta_{\rm F}=1.02\pm0.22$, favours a marginally ($\approx1\sigma$) weaker level of IGM gas infall compared to the most recent results.

\citetalias{2014MNRAS.442.2094T} also compared their result with the simulations of \citet{2003ApJ...585...34M}. Based on several re-simulations varying the simulation parameters, \citet{2003ApJ...585...34M} predict a Ly$\alpha$ infall parameter of $\beta_{\rm F}=1.58\pm0.05$ (or $\beta_{\rm F}=1.45$ taking into account redshift evolution). Although the tension between this and our own results is now smaller, we still find a significantly lower infall parameter than predicted by \citet{2003ApJ...585...34M} at the $\approx3\sigma$ level.

Interestingly, more recent simulations place the predictions of \citet{2003ApJ...585...34M} towards the higher end of published predictions. \citet{2012JCAP...03..004S} noted that the RSD parameter could feasibly lie within the range $0.5\lesssim\beta_{\rm F}\lesssim1.5$ assuming a realistic range of bias parameters for the Ly$\alpha$ forest. \citet{2015JCAP...12..017A} predict the evolution of $\beta$ over the redshift range $2.2<z<3$, finding $\beta_{\rm F}=1.405\pm0.061$ at $z=2.2$ decreasing to $\beta_{\rm F}=1.205\pm0.049$ at $z=3.0$. At the mean redshift of our galaxy survey, $z\approx2.8$, they predict $\beta_{\rm F}=1.284\pm0.052$, a value that is within $\approx1\sigma$ of our observations. Finally, \citet{2015arXiv151104454L} use the Ly$\alpha$ Mass Association Scheme (LyMAS) and predict $\beta_{\rm F}=0.970\pm0.016$ at $z=2.5$, highlighting the range of predictions made via simulations. Once redshift evolution is taken into account, this lies $\approx1\sigma$ below our observations.

\subsubsection{Small scale velocity dispersion}

The total velocity dispersion imprinted upon the 2D cross-correlation function of $\vdisp=240\pm60$~km~s$^{-1}$ comprises three primary components: the measurement uncertainty on the galaxy redshifts; the thermal broadening of the sightline absorption lines; and the intrinsic velocity dispersion between galaxies and Ly$\alpha$ clouds. For the VLT VIMOS galaxy data, the galaxy redshift errors are $\approx350$~km~s$^{-1}$, whilst the Keck LRIS data used for 9 of the sightlines provides marginally more accurate redshifts at $\approx250$~km~s$^{-1}$. From this, it is evident that our result for the velocity dispersion is dominated by the galaxy redshift errors. Indeed the thermal broadening on the absorption lines is $\approx70$~km~s$^{-1}$ \citetalias{2011MNRAS.414...28C}, which if we combine in quadrature with the average of the instrumental errors on the galaxy redshifts gives $\vdisp=310$~km~s$^{-1}$, i.e. $\approx1\sigma$ larger than the best fitting measurement of $\vdisp$ from the $\xi(\sigma,\pi)$ measurement. Combining this estimate of the instrumental plus thermal broadening effects with our results in Fig.~\ref{fig:full2dxcorrfit}, we calculate a $3\sigma$ upper limit on the intrinsic LBG-Ly$\alpha$ velocity dispersion of $\vdisp<220~\kps$.

Ultimately, we require observations with reduced velocity uncertainties to more closely analyse the intrinsic velocity dispersion, however this approximate upper limit still offers some insights and opportunity for comparison with other works at both low and high redshift. \citet{2006MNRAS.367.1251R} measured a large finger-of-god velocity dispersion effect in the cross-correlation between absorbers and galaxies at $z\sim0$, over scales of $\approx400-600$~km~s$^{-1}$. Similarly, \citet{2012ApJ...751...94R} and \citet{2014MNRAS.445..794T} show, using the KBSS data at $z\sim2.3$, elongations extending to $\approx200$~km~s$^{-1}$ along the line of sight. Conversely, \citet{2014MNRAS.437.2017T} claim an upper limit of $\lesssim120$~km~s$^{-1}$ when analysing the cross-correlation of galaxies and absorbers at $0<z<1$.


\subsection{Absorbers tracing the underlying dark matter distribution}

We now consider the relationship between the Ly$\alpha$ forest, the galaxy population and the underlying dark matter distribution. Following the example of \citet{adelberger03} and \citet{2014MNRAS.437.2017T}, we use the Cauchy-Schwarz inequality to evaluate the connection between the LBG population and the Ly$\alpha$ forest. The Cauchy-Schwarz inequality takes the form:

\begin{equation}
\xi_{\rm ag}^2 \leq \xi_{\rm aa}\xi_{\rm gg}
\end{equation}

\noindent where $\xi_{\rm ag}$ is the cross correlation between absorbers and galaxies, $\xi_{\rm aa}$ is the auto-correlation between absorbers, and $\xi_{\rm gg}$ is the galaxy-galaxy auto-correlation. If the two sides of this equation are equal, then it follows that the two populations being evaluated trace the same dark matter structure and the difference in the clustering biases can be used to surmise the relative masses of dark-matter halos that the populations trace within the overall matter structure. On the other hand, if the equality does not hold, i.e. $\xi_{\rm ag}^2/(\xi_{\rm aa}\xi_{\rm gg})<1$, then this can potentially provide insights into the baryonic physics affecting the two populations (assuming that the standard cosmological paradigm is correct). 

\begin{figure}
\centering
\includegraphics[width=\columnwidth]{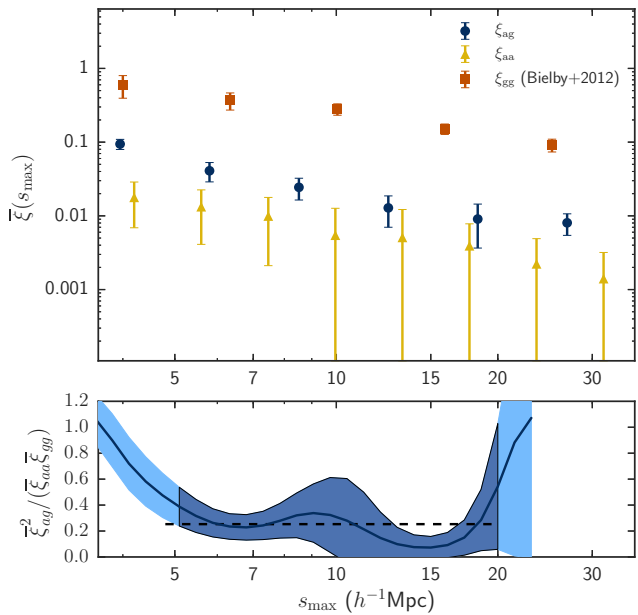}
\caption{\emph{Top panel}: Integrated clustering functions for the LBG-Ly$\alpha$ forest cross correlation; the LBG auto-correlation \citepalias[taken from][]{2013MNRAS.430..425B}; and the Ly$\alpha$ forest auto-correlation. \emph{Lower panel}: The Cauchy-Schwarz ratio calculated from the three correlation functions. The dark shaded region shows the $1\sigma$ range in the ratio over scales of $5\leq s_{\rm max} \leq 20$ representing the linear regime. The dashed line shows the median value in this range.} 
\label{fig:cauchyschwarz}
\end{figure}

We evaluate the Cauchy-Schwarz inequality by calculating the integrated clustering functions, $\overline{\xi}(s_{\rm max})$, of each correlation function, which is given by:

\begin{equation}
\overline{\xi}(s_{\rm max}) = \int^{s_{\rm max}}_0\xi(s)s^2{\rm d}s
\end{equation}

The integrated clustering functions and the Cauchy-Schwarz ratio are shown in Fig.~\ref{fig:cauchyschwarz}. Taking the range $5\leq s_{\rm max}\leq20~h^{-1}$~Mpc (i.e. large enough scales to be in the linear regime, whilst small enough that the uncertainties in the measurements are still relatively low), we find a median value of $\xi_{\rm ag}^2/(\xi_{\rm aa}\xi_{\rm gg})=0.25\pm0.14$. The minimum value in this range is using $s_{\rm max}=12~h^{-1}$~Mpc, which gives $\xi_{\rm ag}^2/(\xi_{\rm aa}\xi_{\rm gg})=0.08\pm0.11$, whilst the maximum is at $s_{\rm max}=20~h^{-1}$~Mpc, which gives $\xi_{\rm ag}^2/(\xi_{\rm aa}\xi_{\rm gg})=0.35\pm0.30$. There is significant variation even in this range then, however the $\xi_{\rm ag}^2 < \xi_{\rm aa}\xi_{\rm gg}$ remains less than unity, maintaining the relationship as an inequality.

That the ratio is less than unity is  a  very  strong  indication that the underlying baryonic matter distributions giving rise to the Ly$\alpha$ forest absorption systems and galaxies are not linearly dependent. This adds to results at low-redshift where weak systems such as populate the forest give $\xi_{\rm ag}^2/(\xi_{\rm aa}\xi_{\rm gg})<1$, whilst stronger absorbers ($N_{\rm HI}\gtrsim10^{14}$~cm$^{-2}$) are consistent with $\xi_{\rm ag}^2/(\xi_{\rm aa}\xi_{\rm gg})=1$ \citep{2014MNRAS.437.2017T}.

\section{Conclusions}

We have used the VLRS galaxy dataset in conjunction with high quality moderate and high resolution background quasar spectra to investigate the relationship between galaxies and the H{\sc i} gas of the IGM. The analysis incorporates \tqso\ $z\gtrsim3$ quasars, 15 of which have been observed with the VLT X-Shooter instrument. Continuum fitting has been performed on the entire sample in order to derive the normalised flux.

We have calculated the Ly$\alpha$ auto-correlation function using our full sample using both line-of-sight and cross line-of-sight ($\sigma\gtrsim6~h^{-1}$~Mpc) data and fit a power-law form, finding a clustering length of $s_0=0.081\pm0.006~h^{-1}$~Mpc and slope of $\gamma=1.09\pm0.04$. 

Using a large spectroscopic sample of LBGs at $z\approx3$, we determine the LBG-Ly$\alpha$ cross-correlation function, $\xi(s)$. As with the Ly$\alpha$ auto-correlation, we fit the data with a power law, finding a clustering length of $s_0=0.27\pm0.14~h^{-1}$~Mpc and slope of $\gamma=1.1\pm0.2$, improving on the accuracy of our analyses presented in \citetalias{2011MNRAS.414...28C} and \citetalias{2014MNRAS.442.2094T}. Further to this, we calculate the LBG-Ly$\alpha$ cross-correlation function as a function of normalised flux, $T$ (a proxy for column density), finding a significant anti-correlation of the resulting clustering lengths with $T$. This shows that higher density H~{\sc i} absorbers are more strongly clustered around galaxies at $z\sim3$, whilst low density absorbers are only weakly clustered with the galaxy population.

Further to the one-dimensional cross-correlation analysis, we calculate the projected and two-dimensional LBG-Ly$\alpha$ cross-correlation functions. Using the projected correlation function, we constrain the real-space correlation length to be $r_0=0.24\pm0.04~h^{-1}$~Mpc (assuming a fixed slope of $\gamma=1.1$ based on the $\xi(s)$ result). Combining this result with our measurement of $\xi(\sigma,\pi)$, we constrain the dynamical properties of the LBG-H{\sc i} density field as probed by the galaxy survey-QSO sightline data. We find $\beta_{\rm F}=1.02\pm0.22$ and $\vdisp=240\pm60$~km~s$^{-1}$. This presents a new and clear detection of the large scale infall of gas towards high density regions within large scale structure. Our measurement of the velocity dispersion between the galaxy and gas components is consistent with the uncertainties on the galaxy redshift measurements, but does give a $3\sigma$ upper limit on the intrinsic LBG-Ly$\alpha$ velocity dispersion of $\vdisp<220~\kps$, similar to our previous result in \citetalias{2014MNRAS.442.2094T}. 

The combination of our auto-correlation and cross-correlation results, along with our previous results for the LBG auto-correlation function, allows us to evaluate the Cauchy-Schwartz inequality, which we find to be significantly below unity ($\xi_{\rm ag}^2/(\xi_{\rm aa}\xi_{\rm gg})=0.25\pm0.14$). Combined with the weak cross-clustering signal, this highlights how Ly$\alpha$ forest absorbers do not linearly follow the density profile traced by the galaxy population.

\section*{Acknowledgments}

Based on observations collected at the European Organisation for Astronomical Research in the Southern Hemisphere under ESO programmes 075.A-0683, 077.A-0612, 079.A-0442, 081.A-0474, 082.A-0494, 085.A-0327, and 087.A-0906. Also based on data obtained with the NOAO Mayall 4m Telescope at Kitt Peak National Observatory, USA (programme ID: 06A-0133) and the NOAO Blanco 4m Telescope at Cerro Tololo Inter-American Observatory, Chile (programme IDs: 03B-0162, 04B-0022). 

RMB, TS and SLM are funded by the STFC research grant ST/L00075X/1. NHMC thanks the Australian Research Council for \textsl{Discovery Project} grant DP130100568 which in part supported this work. DM is supported by the BASAL CATA grant PFB-06 and the Millennium Institute for Astrophysics MAS grant IC120009. LI acknowledges Conicyt grants Basal-CATA PFB–06/2007 and Anillo ACT1417.

This research has made use of the NASA/IPAC Extragalactic Database (NED) which is operated by the Jet Propulsion Laboratory, California Institute of Technology, under contract with NASA.

\bibliographystyle{mnras}
\bibliography{$HOME/Dropbox/lib/rmb}

\label{lastpage}

\end{document}